\documentclass[12pt]{article}

\renewcommand{\thefootnote}{\fnsymbol{footnote}}

\usepackage{amsbsy,amssymb,latexsym,amsfonts,amsmath}
\usepackage{mathrsfs}
\usepackage{graphicx}
\usepackage{cite}
\usepackage{bm}
\usepackage{here}

\numberwithin{equation}{section}
\newcommand{\bel}[1]{\begin{equation}\label{#1}}                     
\newcommand{\bal}[1]{\begin{eqnarray}\label{#1}}                     
\newcommand{\be}{\begin{equation}}
\newcommand{\ee}{\end{equation}}
\newcommand{\im}{\mathrm{i}}
\newcommand{\ex}{\mathrm{e}}
\newcommand{\de}{\mathrm{d}}

\newcommand{\scr}{\scriptstyle}
\newcommand{\qq}{\qquad}

\newcommand{\bea}{\begin{equation}}
\newcommand{\eea}{\end{equation}}

\begin{document}
%
%
\begin{titlepage}
\begin{flushright}
\normalsize
~~~~
OCU-PHYS 455\\
September, 2016 \\
\end{flushright}

\vspace{15pt}

\begin{center}
{\LARGE Cubic constraints for the resolvents of \\ \vspace{5mm}
 the ABJM matrix model and its cousins}
\end{center}

\vspace{23pt}

\begin{center}
{ H. Itoyama$^{a, b}$\footnote{e-mail: itoyama@sci.osaka-cu.ac.jp},
T. Oota$^b$\footnote{e-mail: toota@sci.osaka-cu.ac.jp}, 
Takao Suyama$^b$\footnote{e-mail: suyama@sci.osaka-cu.ac.jp} 
  and  R. Yoshioka$^b$\footnote{e-mail: yoshioka@sci.osaka-cu.ac.jp}  }\\
%
\vspace{18pt}
%

$^a$\it Department of Mathematics and Physics, Graduate School of Science\\
Osaka City University\\
\vspace{5pt}

$^b$\it Osaka City University Advanced Mathematical Institute (OCAMI)

\vspace{5pt}

3-3-138, Sugimoto, Sumiyoshi-ku, Osaka, 558-8585, Japan \\

\end{center}
%
\vspace{20pt}
\begin{center}
Abstract\\
\end{center}
A set of Schwinger-Dyson equations forming constraints for at most 
 three resolvent functions are considered 
 for a class of Chern-Simons matter matrix models 
 with two nodes labelled by a non-vanishing number $n$. 
The two cases $n=2$ and $n= -2$ label respectively the ABJM matrix model, 
which is the hyperbolic lift of the affine $A_1^{(1)}$ quiver matrix model, 
and the lens space matrix model. In the planar limit, 
we derive two cubic loop equations for the two planar resolvents. 
One of these reduces to the quadratic one when $n = \pm 2$.


\vfill

\end{titlepage}

\renewcommand{\thefootnote}{\arabic{footnote}}
\setcounter{footnote}{0}

\section{Introduction}
Chern-Simons-matter (CSm) matrix models 
 \cite{Mari0207,AKMV0211,HY0311,Mari2005,KWY0909,Suya0912,DT0912,MP0912,Suya1008,DMP1103,
 Mari1104,Suya1106,Suya1304,Suya1605} have attracted considerable attentions in 
 recent years in the context of theory 
 on multiple M2 branes and its generalization \cite{ABJM0806,HLLLP0806,ABJ0807}. 
They belong to a class of two matrix (to be denoted by two nodes in this paper) 
 models connected by the measure factor 
 which is attributed to the contributions from the $n$ bi-fundamental multiplets
 upon localization of the CSm action in three dimensions. 
The cases $n=2$, $n=-2$ correspond to the celebrated ABJM matrix model and 
 the lens space matrix model respectively and are well studied 
 \cite{FHM1106,MP1110,PY1207,KMSS1207,HMO1211,AHS1212,HMO1301,HMMO1306}
 mainly by the Fermi-gas approach.
 
This class of matrix models is interesting also from the point of view
 of the $q$-deformation \cite{JS1998,SKAO9507,AKOS9508,AKMOS9604,Kade9604,
 AY0910,AY1004,MMSS1105,BBT1211,NPP1303,IOY1308,IOY2013,Tan1309,NPPT1312,
 Ohku1404,IOY1408,Spod1409,Zenk1412,Y1512,OAF1512,IOY1602,NNZ1605,AKMMMMOZ1604}
  of the Virasoro/W block \cite{DV0909,MMS1001,IO1003,MMM1003,IOYone1008,
  EPSS1110,ZM1110,MMZ1312}
  and of the 2d-4d connection \cite{AGT0906,Wyll0907,MM0908}
 (For more references, see, for example, \cite{IY1507}.): 
 the $n=1$ case provides a hyperbolic lift of 
 the $A_2$ quiver hermitean matrix model 
 that obeys $W_3$ constraints \cite{MMM1991,KMMMP9208,Kost9208,Kost9509} 
 and that produces \cite{IMO0911} 
 the $su(3)$, $N_f = 6$ Witten-Gaiotto curve \cite{Witt9703,Gaio0904}
 while the $n=2$ case provides a hyperbolic lift of 
 the $A_m^{(1)}~(m=1)$ affine quiver matrix \cite{IO1106} 
model that is defined by the incidence matrix  
 of the extended Cartan matrix and whose spacetime interpretation 
 is yet obscure to us. 
 A class of CSm models labelled by $n$  provides a deformation of 
these cases and we study a set of Schwinger-Dyson equations
forming cubic constraints from this generic point of view in this paper. 
 
In the next section, we briefly recall the partition function 
 of the CSm matrix model with two nodes.
In section three, we consider the Schwinger-Dyson equations 
which take the form of the second and the third order constraints for the two resolvents. 
In section four, the planar limit of the equations derived in section three is taken.
We derive a cubic loop equation for each of the symmetric and antisymmetric
combination of the planar loop resolvents.
The remarkable simplicity takes place in those cases $n = \pm2$, where
  one of the two cubic equations reduces to a quadratic one.
  In Appendix A and B, we give some detail of the derivations of the cubic loop equations.

\section{The partition function}

The partition function of the Chern-Simons-matter matrix models with $2$ nodes
is defined by 
\bel{partfn}
Z:= \int \de^{N_1} u \int \de^{N_2} w \, \ex^{S_{\mathrm{eff}}},
\ee
where the ``effective action'' $S_{\mathrm{eff}}$ is given by
\be
\begin{split}
\ex^{S_{\mathrm{eff}}} &= \prod_{1 \leq i<j \leq N_1}
\left( 2 \sinh \frac{u_i - u_j}{2} \right)^2 \prod_{1 \leq a<b \leq N_2}
\left( 2 \sinh \frac{w_a - w_b}{2} \right)^2  \cr
& \times \prod_{i=1}^{N_1} \prod_{a=1}^{N_2} \left( \cosh \frac{u_i - w_a}{2} \right)^{-n} 
\exp\left( - \frac{\kappa_1}{2 g_s} \sum_{i=1}^{N_1} u_i^2
- \frac{\kappa_2}{2 g_s} \sum_{a=1}^{N_2} w_a^2 \right).
\end{split}
\ee
If we set
\be
g_s = \frac{2 \pi \im}{k}, \qq
k_1 = k \, \kappa_1, \qq k_2 = k \, \kappa_2,
\ee
we have
\be
- \frac{\kappa_i}{2 g_s} = \frac{\im k_i}{4\pi}.
\ee
Then the partition function \eqref{partfn} arises from the localization applied to
a supersymmetric $U(N_1)_{k_1} \times U(N_2)_{k_2}$ Chern-Simons
theory with $n$ bi-fundamental hypermultiplets.

In the following, we assume that $n \neq 0$. The average of a function $f(u,w)$ 
with respect to $Z$ is denoted by
\be
\Bigl\langle f(u,w) \Bigr\rangle:= \frac{1}{Z} \int \de^{N_1} u \int \de^{N_2} w
\, f(u,w) \, \ex^{S_{\mathrm{eff}}}.
\ee

\section{Constraints for resolvents}

The resolvents
\bel{resw}
\Bigl\langle \hat{\omega}_i(z) \Bigr\rangle, \qq
(i=1,2)
\ee
play important roles in matrix models. Here
\be\label{resolvent1}
\hat{\omega}_1(z):= g_s \sum_{i=1}^{N_1} \frac{1}{z - \ex^{u_i}}, 
\qq
\hat{\omega}_2(z):= g_s \sum_{a=1}^{N_2} \frac{1}{z- \ex^{w_a}}.
\ee
In this section, we derive second and third order constraints for $\hat{\omega}_i(z)$.
It is known that instead of \eqref{resw},
the resolvents of the following form is natural in the matrix model of Chern-Simons type:
\be
\Bigl\langle \hat{v}_i(z) \Bigr\rangle \qq
(i=1,2),
\ee
where
\be\label{resolvent2}
\hat{v}_1(z):= g_s \sum_{i=1}^{N_1} \frac{z + \ex^{u_i}}{z - \ex^{u_i}}, 
\qq
\hat{v}_2(z):= g_s \sum_{a=1}^{N_2} \frac{z + \ex^{w_a} }{z- \ex^{w_a}}.
\ee
\eqref{resolvent1} and \eqref{resolvent2} are related by 
\be
\hat{v}_i(z) = 2 \, z \, \hat{\omega}_i(z) - t_i.
\ee
Here $t_i:= N_i g_s$ are the 't Hooft couplings.

For the sake of simplicity, we use $\hat{\omega}_i(z)$ in order to derive constraints for the resolvents.
The constraints for $\hat{\omega}_i(z)$ are easily converted into those for $\hat{v}_i(z)$.

\subsection{Second order constraints}

From
\be
 \int \de^{N_1} u \int\de^{N_2} w
\sum_{i=1}^{N_1} \frac{\partial}{\partial u_i}
\left( \frac{1}{z- \ex^{u_i}}  \, \ex^{S_{\mathrm{eff}}} \right) =  0,
\ee
which is the hyperbolic counterpart of the Virasoro constraints 
\cite{Davi1990,MM1990,IM1991,IM1991W,Itoy9111}, 
we obtain the following constraint:
\be
\begin{split}
&   \left\langle \sum_i \frac{\ex^{u_i}}{(z-\ex^{u_i})^2} \right\rangle 
+ \left\langle \sum_i \sum_{j \neq i}
\frac{1}{(z-\ex^{u_i})} 
\coth \frac{u_i-u_j}{2} 
\right\rangle \cr
&- \frac{n}{2}  \left\langle \sum_{i} \sum_a  
\frac{1}{(z-\ex^{u_i})} \tanh \frac{u_i-w_a}{2} 
\right\rangle
- \frac{\kappa_1}{g_s} \left\langle \sum_i \frac{u_i}{z-\ex^{u_i}}
\right\rangle = 0.
\end{split}
\ee
Using an identity
\be
\sum_i \frac{\ex^{u_i}}{(z-\ex^{u_i})^2}
+  \sum_i \sum_{j \neq i}
\frac{1}{(z-\ex^{u_i})} \coth \frac{u_i-u_j}{2}
= z \left( \sum_i \frac{1}{z-\ex^{u_i}} \right)^2 - N_1 \sum_i \frac{1}{z - \ex^{u_i}},
\ee
we can rewrite the above constraint as follows
\bel{LEQ1}
z \bigl\langle \hat{\omega}_1(z)^2 \bigr\rangle 
- \bigl( t_1 + \kappa_1 \log z \bigr) \bigl\langle \hat{\omega}_1(z) \rangle 
= \frac{n}{2}  \left\langle \widehat{R}^{(2)}_1(z)
\right\rangle
-   \left\langle \widehat{F}_1(z)
\right\rangle,
\ee
where
\be
\widehat{F}_1(z):= \kappa_1 g_s \sum_{i=1}^{N_1} \frac{\log z - u_i}{z- \ex^{u_i}},
\ee
\be
\widehat{R}^{(2)}_1(z):= g_s^2 \sum_{i=1}^{N_1} \sum_{a=1}^{N_2}
\frac{1}{z-\ex^{u_i}} \tanh \frac{u_i - w_a}{2}.
\ee
Here $\log z$ takes a real value on the positive $\mathrm{Re}\, z$ axis 
and has a cut along the negative $\mathrm{Re}\, z$ axis.

Similarly, from
\be
\int \de^{N_1} u \int \de^{N_2} w \sum_{a=1}^{N_2} \frac{\partial}{\partial w_a}
\left( \frac{1}{z + \ex^{w_a}} \, \ex^{S_{\mathrm{eff}}} \right) =0,
\ee
we obtain 
\bel{LEQ4}
 z \bigl\langle \hat{\omega}_2(-z)^2 \bigr\rangle 
+  \bigl( t_2 + \kappa_2 \log(-z)  \bigr)
\bigl\langle \hat{\omega}_2(-z) \bigr\rangle 
= - \frac{n}{2}  \left\langle \widehat{R}^{(2)}_2(-z)
\right\rangle
+   \left\langle \widehat{F}_2(-z)
\right\rangle,
\ee
where
\be
\widehat{F}_2(z):= \kappa_2 g_s \sum_{a=1}^{N_2} \frac{\log z - w_a}{z- \ex^{w_a}},
\ee
\be
\widehat{R}^{(2)}_2(z):= g_s^2 \sum_{i=1}^{N_1} \sum_{a=1}^{N_2}
\frac{1}{z- \ex^{w_a}} \tanh \frac{w_a - u_i}{2}.
\ee
Here  $\log(-z)$ has a cut along the positive $\mathrm{Re}\, z$ axis and takes
real values on the negative $\mathrm{Re}\, z$ axis.

By adding \eqref{LEQ1} and \eqref{LEQ4}, we find the second oder constraints for
the resolvent operators $\hat{\omega}_i(z)$:
\bel{VC1}
\begin{split}
&z \Bigl\langle \hat{\omega}_1(z)^2 + n \, \hat{\omega}_1(z) \hat{\omega}_2(-z)
+ \hat{\omega}_2(-z)^2 \Bigr\rangle \cr
&- \left( t_1 - \frac{n}{2} t_2 + \kappa_1 \log z \right) \langle \hat{\omega}_1(z) \rangle
+ \left( t_2 - \frac{n}{2} t_1 + \kappa_2 \log (-z) \right) \langle \hat{\omega}_2(-z) \rangle \cr
&=-   \left\langle \widehat{F}_1(z)
\right\rangle
+ \left\langle \widehat{F}_2(-z)
\right\rangle.
\end{split}
\ee
Here we  have used the following identity:
\bel{R2id}
\begin{split}
\widehat{R}^{(2)}_1(z) - \widehat{R}^{(2)}_2(-z) 
&= g_s^2 \sum_i \sum_a \left(\frac{1}{z-\ex^{u_i}} - \frac{1}{z+\ex^{w_a}} \right)
\tanh \frac{u_i - w_a}{2} \cr
&=g_s^2 \sum_i \sum_a 
\left( \frac{2z}{(z-\ex^{u_i})(z+\ex^{w_a})} - \frac{1}{z - \ex^{u_i}} - \frac{1}{z+\ex^{w_a}}
\right) \cr
&= - 2 z \, \hat{\omega}_1(z)\, \hat{\omega}_2(-z) - t_2 \, \hat{\omega}_1(z)
+ t_1 \, \hat{\omega}_2(-z).
\end{split}
\ee

\subsection{Third order constraints}

There are four third order constraints. 

\subsubsection{Third order constraint 1}

From
\bel{TOC1}
\int \de^{N_1} u \int \de^{N_2} w \sum_{i=1}^{N_1} \sum_{\stackrel{\scr j=1}{(j \neq i)}}^{N_1}
\frac{\partial}{\partial u_i} \left(
\frac{2}{z - \ex^{u_i}} \coth \frac{u_i - u_j}{2}\,  \ex^{S_{\mathrm{eff}}} \right) = 0,
\ee
we obtain a constraint
\bel{HLeq1}
S_{11}^{(0)}(z) + g_s \, S_{11}^{(1)}(z) = 0,
\ee
where
\bel{S110F}
\begin{split}
S_{11}^{(0)}(z)
&= \frac{8z^2}{3} \Bigl\langle \hat{\omega}_1(z)^3 \Bigr\rangle
- 2 z ( 2 t_1 - 4 g_s + \kappa_1 \log z )
\Bigl\langle \hat{\omega}_1(z)^2 \Bigr\rangle \cr
&+ 2 (t_1 - g_s) (t_1 - 2 g_s +\kappa_1 \log z) \Bigl\langle \hat{\omega}_1(z)
\Bigr\rangle - n \left\langle \widehat{R}^{(3)}_1(z) \right\rangle 
+ 2 \left\langle \widehat{H}_1(z)
\right\rangle \cr
& + 4 z^2 g_s \Bigl\langle \hat{\omega}_1(z)^2 \Bigr\rangle'
-2z (2t_1 - 4 g_s +\kappa_1 \log z ) g_s \Bigl\langle \hat{\omega}_1(z) \Bigr\rangle' 
+ \frac{8z^2}{3} g_s^2 \Bigl\langle \hat{\omega}_1(z) \Bigr\rangle'',
\end{split}
\ee
\bel{S111F}
\begin{split}
S_{11}^{(1)}(z) &=\left( - 2 z^2 \bigl\langle \hat{\omega}_1(z)^2 \bigr\rangle
+  2(t_1 -g_s ) z  \bigl\langle \hat{\omega}_1(z) \bigr\rangle
- 2z^2 g_s  \bigl\langle \hat{\omega}_1(z) \bigr\rangle'  \right)' \cr
& + 2 (t_1 - g_s ) \Bigl\langle \hat{\omega}_1(z) \Bigr\rangle
+ g_s^2
\left\langle \sum_i \sum_{j \neq i}
\frac{1}{(z-\ex^{u_i})} \frac{1}{ \sinh^2 \frac{u_i-u_j}{2} }
\right\rangle.
\end{split}
\ee
Here ${}'$ denotes the derivative with respect to $z$ and
\be
\widehat{R}^{(3)}_1(z):= g_s^3 \sum_i \sum_a \sum_{j \neq i}
\frac{1}{z-\ex^{u_i}}\, \tanh \frac{u_i-w_a}{2} \coth \frac{u_i - u_j}{2},
\ee
\be
\widehat{H}_1(z):= \kappa_1 g_s^2 \sum_{i} \sum_{j \neq i}
\frac{\log z - u_i}{z- \ex^{u_i}} \coth \frac{u_i - u_j}{2}.
\ee
See Appendix \ref{A} for details.

Since we have assumed that $n \neq 0$, 
the third order constraint \eqref{HLeq1} is equivalent to the following condition
\bel{R13F}
\begin{split}
\left\langle \widehat{R}_1^{(3)}(z) \right\rangle
&=\frac{8z^2}{3n} \Bigl\langle \hat{\omega}_1(z)^3 \Bigr\rangle
- \frac{2 z}{n} ( 2 t_1 - 4 g_s + \kappa_1 \log z )
\Bigl\langle \hat{\omega}_1(z)^2 \Bigr\rangle \cr
&+ \frac{2}{n} (t_1 - g_s) (t_1 - 2 g_s +\kappa_1 \log z) \Bigl\langle \hat{\omega}_1(z)
\Bigr\rangle 
 + \frac{2}{n}  \left\langle \widehat{H}_1(z)
\right\rangle \cr
& + \frac{4 z^2}{n} g_s \Bigl\langle \hat{\omega}_1(z)^2 \Bigr\rangle'
-\frac{2z}{n} (2t_1 - 4 g_s +\kappa_1 \log z ) g_s \Bigl\langle \hat{\omega}_1(z) \Bigr\rangle' \cr
& + \frac{8z^2}{3n} g_s^2 \Bigl\langle \hat{\omega}_1(z) \Bigr\rangle''
+ \frac{g_s}{n} S_{11}^{(1)}(z).
\end{split}
\ee

\subsubsection{Third order constraint 2}

Similar to the case of constraint 1, 
\bel{TOC2}
\int \de^{N_1} u \int \de^{N_2} w \sum_{a=1}^{N_2} 
\sum_{\stackrel{\scr b=1}{(b \neq a)}}^{N_2} 
\frac{\partial}{\partial w_a} \left( \frac{2 }{z + \ex^{w_a}}
\coth \frac{w_a-w_b}{2} \, \ex^{S_{\mathrm{eff}}} \right) = 0
\ee
implies the following condition
\bel{R23F}
\begin{split}
\left\langle \widehat{R}_2^{(3)}(-z) \right\rangle
&=\frac{8z^2}{3n} \Bigl\langle \hat{\omega}_2(-z)^3 \Bigr\rangle
+ \frac{2 z}{n} ( 2 t_2 - 4 g_s + \kappa_2 \log (-z) )
\Bigl\langle \hat{\omega}_2(-z)^2 \Bigr\rangle \cr
&+ \frac{2}{n} (t_2 - g_s) (t_2 - 2 g_s +\kappa_2 \log (-z)) \Bigl\langle \hat{\omega}_2(-z)
\Bigr\rangle 
+ \frac{2}{n} \left\langle \widehat{H}_2(-z) \right\rangle \cr
& - \frac{4 z^2}{n} g_s \Bigl\langle \hat{\omega}_2(-z)^2 \Bigr\rangle'
-\frac{2z}{n} (2t_2 - 4 g_s +\kappa_2 \log (-z) ) g_s \Bigl\langle \hat{\omega}_2(-z) \Bigr\rangle' \cr
& + \frac{8z^2}{3n} g_s^2 \Bigl\langle \hat{\omega}_2(-z) \Bigr\rangle''
+ \frac{g_s}{n} S_{22}^{(1)}(-z),
\end{split}
\ee
where
\be
\begin{split}
S_{22}^{(1)}(z) &:=\left( - 2 z^2 \bigl\langle \hat{\omega}_2(z)^2 \bigr\rangle
+  2(t_2 -g_s ) z  \bigl\langle \hat{\omega}_2(z) \bigr\rangle
- 2z^2 g_s  \bigl\langle \hat{\omega}_2(z) \bigr\rangle'  \right)' \cr
& + 2 (t_2 - g_s ) \Bigl\langle \hat{\omega}_2(z) \Bigr\rangle
+ g_s^2
\left\langle \sum_a \sum_{b \neq a}
\frac{1}{(z-\ex^{w_a})} \frac{1}{ \sinh^2 \frac{w_a-w_b}{2} }
\right\rangle.
\end{split}
\ee
Here 
\be
\widehat{R}^{(3)}_2(z):= g_s^3 \sum_i \sum_a \sum_{b \neq a}
\frac{1}{z-\ex^{w_a}}\, \tanh \frac{w_a-u_i}{2} \coth \frac{w_a - w_b}{2},
\ee
\be
\widehat{H}_2(z):= \kappa_2 g_s^2 \sum_a \sum_{b \neq a}
\frac{\log z - w_a}{z- \ex^{w_a}} \coth \frac{w_a-w_b}{2}.
\ee
In our convention, 
${}'$ always denotes the derivative with respect to $z$. Hence under $z \rightarrow -z$, 
it transforms as follows:
\be
f(z) ' = \frac{\partial}{\partial z} f(z) \longrightarrow \frac{\partial}{\partial (-z)} f(-z)
= - \frac{\partial}{\partial z} f(-z) = 
-\bigl( f(-z) \bigr)'.
\ee

\subsubsection{Third order constraint 3}

From
\bel{TOC3}
\int \de^{N_1} u \int \de^{N_2} w
\sum_{i=1}^{N_1} \sum_{a=1}^{N_2} \frac{\partial}{\partial u_i}
\left( \frac{2}{z - \ex^{u_i}} \tanh \frac{u_i - w_a}{2} \, \ex^{S_{\mathrm{eff}}}
\right) =0,
\ee
we obtain a constraint
\bel{HLeq3}
S_{12}^{(0)}(z) + g_s\, S_{12}^{(1)}(z) = 0,
\ee
where 
\bel{S120F}
\begin{split}
S^{(0)}_{12}(z) &= 2  \left\langle \widehat{R}^{(3)}_1(z) \right\rangle
 - 2 n \left\langle \widehat{R}^{(3)}_2(-z) \right\rangle \cr
 & - n \Bigl\langle \hat{\omega}_1(z) \bigl( 2 z \, \hat{\omega}_2(-z) + t_2 \bigr)^2
\Bigr\rangle 
 + 2 n  t_1 z \Bigl\langle \hat{\omega}_2(-z)^2 \Bigr\rangle
+ 2n  t_1 (t_2-g_s) \Bigl\langle \hat{\omega}_2(-z) \Bigr\rangle \cr
& - 2 \kappa_1  \log z\, \left\langle 
\widehat{R}_1^{(2)}(z)
\right\rangle 
+ 2  \left\langle \widehat{G}_1(z) \right\rangle \cr
&+ 4n z^2 g_s  \Bigl\langle \hat{\omega}_1(z) \bigl( \hat{\omega}_2(-z)  \bigr)' \Bigr\rangle
+ 4n z g_s \Bigl\langle \hat{\omega}_1(z) \, \hat{\omega}_2(-z) \Bigr\rangle \cr
&- 2 n t_1 z  g_s \Bigl\langle \hat{\omega}_2(-z) \Bigr\rangle' 
+ n  t_2 g_s\Bigl\langle \hat{\omega}_1(z) \Bigr\rangle,
\end{split}
\ee
\bel{S121F}
S_{12}^{(1)}(z) = - n t_2 \Bigl\langle \hat{\omega}_1(z) \Bigr\rangle
-2 \left(  z \left\langle \widehat{R}_1^{(2)}(z) \right\rangle \right)' 
+ (n+1) g_s^2 
\left\langle \sum_i \sum_a \frac{1}{z-\ex^{u_i}} \frac{1}{\cosh^2 \frac{u_i - w_a}{2}}
\right\rangle.
\ee
Here
\be
\widehat{G}_1(z):= \kappa_1 g_s^2 \sum_i \sum_a
\frac{\log z - u_i}{z - \ex^{u_i}} \tanh \frac{u_i - w_a}{2}.
\ee
See Appendix \ref{B} for details.

Using the third order constraints \eqref{R13F}, \eqref{R23F} and the second order constraint
\eqref{LEQ1}, we can rewrite the terms containing $\langle \widehat{R}_1^{(3)}(z) \rangle$,
 $\langle \widehat{R}_2^{(3)}(-z) \rangle$ and
 $\langle \widehat{R}_1^{(2)}(z) \rangle$.

\subsubsection{Third order constraint 4}

Similar to the case of the third order constraint 3, 
\bel{TOC4}
\int \de^{N_1} u \int \de^{N_2} w \sum_{i=1}^{N_1} \sum_{a=1}^{N_2}
\frac{\partial}{\partial w_a}
\left( \frac{2}{z + \ex^{w_a}} \tanh \frac{w_a-u_i}{2} \, \ex^{S_{\mathrm{eff}}}
\right) = 0
\ee
leads to
\be
S_{21}^{(0)}(-z) + g_s S_{21}^{(1)}(-z) = 0,
\ee
where
\be
\begin{split}
S^{(0)}_{21}(-z) &= 2  \left\langle \widehat{R}^{(3)}_2(-z) \right\rangle
 - 2 n \left\langle \widehat{R}^{(3)}_1(z) \right\rangle \cr
 & - n \Bigl\langle \hat{\omega}_2(-z) \bigl( 2 z \, \hat{\omega}_1(z) - t_1 \bigr)^2
\Bigr\rangle 
 - 2 n  t_2 z \Bigl\langle \hat{\omega}_1(z)^2 \Bigr\rangle
+ 2n  t_2 (t_1-g_s) \Bigl\langle \hat{\omega}_1(z) \Bigr\rangle \cr
& - 2 \kappa_2  \log (-z)\, \left\langle 
\widehat{R}_2^{(2)}(-z)
\right\rangle 
+ 2  \left\langle \widehat{G}_2(-z) \right\rangle \cr
&- 4n z^2 g_s  \Bigl\langle \hat{\omega}_2(-z) \bigl( \hat{\omega}_1(z)  \bigr)' \Bigr\rangle
- 4n z g_s \Bigl\langle \hat{\omega}_2(-z) \, \hat{\omega}_1(z) \Bigr\rangle \cr
&- 2 n t_2 z  g_s \Bigl\langle \hat{\omega}_1(z) \Bigr\rangle' 
+ n  t_1 g_s\Bigl\langle \hat{\omega}_2(-z) \Bigr\rangle,
\end{split}
\ee 
\be
\begin{split}
S_{21}^{(1)}(-z) &= - n t_1 \Bigl\langle \hat{\omega}_2(-z) \Bigr\rangle
-2 \left(  z \left\langle \widehat{R}_2^{(2)}(-z) \right\rangle \right)'  \cr
& - (n+1) g_s^2 
\left\langle \sum_i \sum_a \frac{1}{z+\ex^{w_a}} \frac{1}{\cosh^2 \frac{w_a - u_i}{2}}
\right\rangle.
\end{split}
\ee
Here
\be
\widehat{G}_2(z):= \kappa_2 g_s^2 \sum_i \sum_a
\frac{\log z - w_a}{z- \ex^{w_a}} \tanh \frac{w_a-u_i}{2}.
\ee

\subsubsection{Summary: third order constraints}

We have obtained the following third order constraints:
\bel{FHC1}
\begin{split}
& 2  \left\langle \widehat{R}^{(3)}_1(z) \right\rangle
 - 2 n \left\langle \widehat{R}^{(3)}_2(-z) \right\rangle \cr
 & - n \Bigl\langle \hat{\omega}_1(z) \bigl( 2 z \, \hat{\omega}_2(-z) + t_2 \bigr)^2
\Bigr\rangle 
 + 2 n  t_1 z \Bigl\langle \hat{\omega}_2(-z)^2 \Bigr\rangle
+ 2n  t_1 (t_2-g_s) \Bigl\langle \hat{\omega}_2(-z) \Bigr\rangle \cr
& - 2 \kappa_1  \log z\, \left\langle 
\widehat{R}_1^{(2)}(z)
\right\rangle 
+ 2  \left\langle \widehat{G}_1(z) \right\rangle \cr
&+ 4n z^2 g_s  \Bigl\langle \hat{\omega}_1(z) \bigl( \hat{\omega}_2(-z)  \bigr)' \Bigr\rangle
+ 4n z g_s \Bigl\langle \hat{\omega}_1(z) \, \hat{\omega}_2(-z) \Bigr\rangle 
- 2 n t_1 z  g_s \Bigl\langle \hat{\omega}_2(-z) \Bigr\rangle' \cr 
&- 2 g_s  \left(  z \left\langle \widehat{R}_1^{(2)}(z) \right\rangle \right)' 
+ (n+1) g_s^3 
\left\langle \sum_i \sum_a \frac{1}{z-\ex^{u_i}} \frac{1}{\cosh^2 \frac{u_i - w_a}{2}}
\right\rangle=0,
\end{split}
\ee
\bel{FHC2}
\begin{split}
& 2  \left\langle \widehat{R}^{(3)}_2(-z) \right\rangle
 - 2 n \left\langle \widehat{R}^{(3)}_1(z) \right\rangle \cr
 & - n \Bigl\langle \hat{\omega}_2(-z) \bigl( 2 z \, \hat{\omega}_1(z) - t_1 \bigr)^2
\Bigr\rangle 
 - 2 n  t_2 z \Bigl\langle \hat{\omega}_1(z)^2 \Bigr\rangle
+ 2n  t_2 (t_1-g_s) \Bigl\langle \hat{\omega}_1(z) \Bigr\rangle \cr
& - 2 \kappa_2  \log (-z)\, \left\langle 
\widehat{R}_2^{(2)}(-z)
\right\rangle 
+ 2  \left\langle \widehat{G}_2(-z) \right\rangle \cr
&- 4n z^2 g_s  \Bigl\langle \hat{\omega}_2(-z) \bigl( \hat{\omega}_1(z)  \bigr)' \Bigr\rangle
- 4n z g_s \Bigl\langle \hat{\omega}_2(-z) \, \hat{\omega}_1(z) \Bigr\rangle 
- 2 n t_2 z  g_s \Bigl\langle \hat{\omega}_1(z) \Bigr\rangle' \cr
&-2 g_s \left(  z \left\langle \widehat{R}_2^{(2)}(-z) \right\rangle \right)' 
- (n+1) g_s^3 
\left\langle \sum_i \sum_a \frac{1}{z+\ex^{w_a}} \frac{1}{\cosh^2 \frac{w_a - u_i}{2}}
\right\rangle = 0,
\end{split}
\ee
where $\langle \widehat{R}_1^{(3)}(z) \rangle$ 
and $\langle \widehat{R}_2^{(3)}(-z) \rangle$ are 
respectively given by \eqref{R13F} and \eqref{R23F}.
The second order constraints also imply
\be
\left\langle \widehat{R}_1^{(2)}(z) \right\rangle
=\frac{2z}{n} \bigl\langle \hat{\omega}_1(z)^2 \bigr\rangle
- \frac{2}{n} ( t_1 + \kappa_1 \log z) 
\bigl\langle \hat{\omega}_1(z) \bigr\rangle 
+ \frac{2}{n} \left\langle \widehat{F}_1(z)
\right\rangle,
\ee
\be
\left\langle \widehat{R}_2^{(2)}(-z) \right\rangle
=-\frac{2z}{n} \bigl\langle \hat{\omega}_2(-z)^2 \bigr\rangle
- \frac{2}{n} ( t_2 + \kappa_2 \log (-z) ) 
\bigl\langle \hat{\omega}_2(-z) \bigr\rangle 
+ \frac{2}{n} \left\langle \widehat{F}_2(-z)
\right\rangle.
\ee

\section{Planar limit and loop equations}

Keeping the 't Hooft couplings $t_1=N_1 g_s$ and $t_2= N_2 g_s$ finite, we take 
the planar limit $g_s \rightarrow 0$.

For $i=1,2$, let
\be
\omega_{i}(z) := \lim_{g_s \rightarrow 0} \bigl\langle \hat{\omega}_i(z) \bigr\rangle,
\qq
f_i(z) := \lim_{g_s \rightarrow 0} \left\langle \widehat{F}_i(z) \right\rangle,
\ee
\be
g_i(z) := \lim_{g_s \rightarrow 0} \left\langle \widehat{G}_i(z) \right\rangle,
\qq
h_i(z) := \lim_{g_s \rightarrow 0} \left\langle \widehat{H}_i(z) \right\rangle,
\ee
\be
r_i^{(2)}(z) := \lim_{g_s \rightarrow 0} \left\langle \widehat{R}_i^{(2)}(z) \right\rangle,
\qq
r_i^{(3)}(z) := \lim_{g_s \rightarrow 0} \left\langle \widehat{R}_i^{(3)}(z) \right\rangle,
\ee
We also introduce
\be
v_i(z):= \lim_{g_s \rightarrow 0} \bigl\langle \hat{v}_i(z) \bigr\rangle = 2\, z \, \omega_i(z) - t_i.
\ee

\subsection{Planar second order constraint}

The planar limit of the second order constraint \eqref{VC1} is given by
\bel{PVC}
\begin{split}
& z \omega_1(z)^2 + nz \omega_1(z) \omega_2(-z) + z \omega_2(-z)^2 \cr
& - A_1(z)  \omega_1(z)
+ A_2(-z) \omega_2(-z) + f_1(z) - f_2(-z) = 0,
\end{split}
\ee
where
\be
A_1(z):= t_1 - \frac{n}{2} t_2 + \kappa_1 \log z, \qq
A_2(z):= t_2 - \frac{n}{2} t_1 + \kappa_2 \log z.
\ee

\subsection{Planar third order constraints}

The planar limit of the third order constraints \eqref{FHC1} and \eqref{FHC2} are 
 respectively given by
\bel{PFHC1}
\begin{split}
& 2 r_1^{(3)}(z) - 2 n r_2^{(3)}(-z) - n \omega_1(z)( 2 z \omega_2(-z)+t_2)^2 \cr
& + 2n t_1 z \omega_2(-z)^2 + 2n t_1 t_2 \, \omega_2(-z)
- 2\kappa_1 \log z \, r_1^{(2)}(z) + 2 g_1(z) = 0,
\end{split}
\ee
\bel{PFHC2}
\begin{split}
& 2 r_2^{(3)}(-z) - 2n r_1^{(3)}(z)
-  n \omega_2(-z) ( 2 z \omega_1(z) - t_1)^2 \cr
& - 2n t_2 z \omega_1(z)^2 + 2n t_1 t_2 \, \omega_1(z)
- 2 \kappa_2 \log(-z)\, r_2^{(2)}(-z) + 2 g_2(-z) = 0,
\end{split}
\ee
where
\be
\begin{split}
r_1^{(3)}(z)
&= \frac{8z^2}{3n} \omega_1(z)^3 - \frac{2z}{n} ( 2 t_1 + \kappa_1 \log z) \omega_1(z)^2 \cr
&+ \frac{2}{n} t_1 (t_1 + \kappa_1 \log z) \omega_1(z) + \frac{2}{n} h_1(z), \cr
r_2^{(3)}(-z)
&= \frac{8z^2}{3n} \omega_2(-z)^3 + \frac{2z}{n} ( 2 t_2 + \kappa_2\log(- z))\,  \omega_2(-z)^2 \cr
&+ \frac{2}{n} t_2 (t_2 + \kappa_2 \log(- z)) \omega_2(-z) + \frac{2}{n} h_2(-z), \cr
r_1^{(2)}(z) &= \frac{2z}{n} \omega_1(z)^2 - \frac{2}{n} ( t_1 + \kappa_1 \log z)
\omega_1(z) + \frac{2}{n} f_1(z),
\cr
r_2^{(2)}(-z) &= -\frac{2z}{n} \omega_2(-z)^2 - \frac{2}{n} ( t_2 + \kappa_2 \log (-z))
\omega_2(-z) + \frac{2}{n} f_2(-z).
\end{split}
\ee

\subsubsection{Explicit forms}

The explicit form of 
the planar third order constraints \eqref{PFHC1} and \eqref{PFHC2} are respectively 
\bel{PHC1}
\begin{split}
& \frac{16z^2}{3n} \omega_1(z)^3 - 4n z^2 \omega_1(z) \omega_2(-z)^2
- \frac{16z^2}{3} \omega_2(-z)^3 \cr
& - \frac{8z}{n} (t_1 + \kappa_1 \log z) \omega_1(z)^2 
- 4n t_2 z \, \omega_1(z) \omega_2(-z)
 - 2z \bigl( 4 t_2 - n t_1 + 2\kappa_2 \log(-z)  \bigr) \omega_2(-z)^2 \cr
& + \frac{1}{n}( 2 t_1 - n t_2 + 2 \kappa_1 \log z )( 2 t_1 + n t_2 + 2\kappa_1 \log z)
\omega_1(z) \cr
&- 2 t_2 (2 t_2 - n t_1 + 2\kappa_1 \log(-z))
\omega_2(-z) \cr
& + \frac{4}{n} h_1(z) - \frac{4\kappa_1}{n} \log z\, f_1(z)
+ 2 g_1(z) - 4 h_2(-z) = 0,
\end{split}
\ee
\bel{PHC2}
\begin{split}
& - \frac{16 z^2}{3} \omega_1(z)^3 - 4nz^2 \omega_1(z)^2 \omega_2(-z)
+ \frac{16z^2}{3n} \omega_2(-z)^3 \cr
& +  2z( 4 t_1 + 2 \kappa_1 \log z - n t_2) \omega_1(z)^2 
+ 4 n t_1 z \omega_1(z) \omega_2(-z) + \frac{8 z}{n} (t_2 + \kappa_2 \log(-z)) \omega_2(-z)^2 \cr
& - 2 t_1 ( 2 t_1 + 2\kappa_1 \log z - n t_2) \omega_1(z) \cr
& + \frac{1}{n} ( 2 t_2 - n t_1 + 2\kappa_2 \log(-z))(2 t_2 + n t_1 + 2\kappa_2 \log(-z))
\omega_2(-z) \cr
& + \frac{4}{n} h_2(-z) - \frac{4\kappa_2}{n} \log(-z)\, f_2(-z) + 2 g_2(-z) - 4 h_1(z)  = 0.
\end{split}
\ee

\subsection{Cubic loop equations}

Let
\be
\omega_{\pm}(z):= \omega_1(z) \pm \omega_2(-z),
\qq
f_{\pm}(z):= f_1(z) \pm f_2(-z),
\ee
\be
g_{\pm}(z):= g_1(z) \pm g_2(-z), 
\qq
h_{\pm}(z):= h_1(z) \pm h_2(-z),
\ee
\be
v_{\pm}(z):= v_1(z) \pm v_2(-z),
\ee
\be
A_{\pm}(z):= A_1(z) \pm A_2(-z),
\qq
t_{\pm}:= t_1 \pm t_2.
\ee
Note that
\be\label{A_pm}
A_+(z) = - \frac{(n-2)}{2} t_+ + K_+(z),
\qq
A_-(z) = \frac{(n+2)}{2} t_- + K_-(z),
\ee
where
\be\label{K_pm}
K_{\pm}(z):= \kappa_1 \log z \pm \kappa_2 \log(-z).
\ee
In terms of $\omega_{\pm}(z)$, the planar second order constraint \eqref{PVC} can be rewritten as
\bel{PVCr}
\frac{(n+2)}{2} z \, \omega_+(z)^2 - \frac{(n-2)}{2} z \, \omega_-(z)^2   
-  A_-(z) \omega_+(z) -  A_+(z) \omega_-(z)
+ 2 f_-(z) = 0.
\ee

\subsubsection{Cubic equation for $\omega_+(z)$}

By adding \eqref{PHC1} and \eqref{PHC2},  we find
\be
\begin{split}
& - \frac{(n+2)(3n-2)}{3n} z^2 \, \omega_+(z)^3
+ \left\{ (n+2) t_-  + \frac{(n-2)}{n} A_-(z) \right\} z \, \omega_+(z)^2 \cr
& + \frac{1}{n} \Bigl( A_+(z)^2 + A_-(z)^2 - 2 n t_- A_-(z) \Bigr) \omega_+(z) \cr
&+\frac{1}{n} \Bigl\{ (n-2) z \omega_-(z)^2 + 2 A_+(z) \omega_-(z) \Bigr\}
\Bigl\{ (n-2)z \, \omega_+(z) - n t_- + A_-(z) \Bigr\} \cr
& - \frac{2}{n} K_+(z)  f_+(z)
- \frac{2}{n} K_-(z)  f_-(z) - \frac{4(n-1)}{n} h_+(z)  + 2 g_+(z)= 0.
\end{split}
\ee
Using the planar second order constraint \eqref{PVCr}, we can convert this equation into
an algebraic equation for $\omega_+(z)$:
\bel{cubic1}
\begin{split}
& - \frac{4(n+2)}{3} z^2\, \omega_+(z)^3
+ 4 A_-(z) z\, \omega_+(z)^2 \cr
&+ \Bigl\{ A_+(z)^2 - A_-(z)^2 + 4 (n-2) z f_-(z)  \Bigr\} \omega_+(z) \cr
& -  2 K_+(z)   f_+(z) 
-  2 \Bigl\{ (n-2)  t_- - K_-(z) \Bigr\} f_-(z) \cr
& - 4(n-1) h_+(z) + 2 n  g_+(z)  = 0.
\end{split}
\ee
For $n \neq -2$, this is a cubic equation. For $n=-2$, it is quadratic.

When $n \neq -2$, in terms of
\be
x_-(z):= (n+2)  z \, \omega_+(z) - A_-(z)
= \frac{(n+2)}{2} \, v_-(z) - K_-(z),
\ee
the cubic equation \eqref{cubic1} becomes
\bel{cubic1r}
x_-(z)^3 - 3 \, p(z) \, x_-(z) - q_+(z) = 0,
\ee
where
\bel{pz}
p(z) = \frac{1}{4}
\Bigl\{ (n+2) A_+(z)^2 - (n-2) A_-(z)^2 + 4 (n+2)(n-2)z f_-(z) \Bigr\},
\ee
\be
\begin{split}
q_+(z) &= \frac{1}{4} A_-(z) \Bigl\{ 3(n+2) A_+(z)^2
- (3n-2) A_-(z)^2 \Bigr\} \cr
&- \frac{3(n+2)^2}{2}  z\, K_+(z)  f_+(z) 
+ \frac{3(n+2)(3n-2)}{2} z\, K_-(z)   f_-(z) \cr
&- 3(n+2)^2 (n-1) z\,  h_+(z) + \frac{3 n(n+2)^2}{2} z \, g_+(z).
\end{split}
\ee

\subsubsection{Cubic equation for $\omega_-(z)$}

By subtracting \eqref{PHC2} from \eqref{PHC1}, we obtain
\be
\begin{split}
& - \frac{(3n+2)(n-2)}{3n} z^2\, \omega_-(z)^3
+ \left\{ (n-2) t_+ - \frac{(n+2)}{n} A_+(z) \right\} z \, \omega_-(z)^2 \cr
&+\frac{1}{n} \Bigl( A_+(z)^2 + A_-(z)^2 + 2n t_+ A_+(z) \Bigr) \omega_-(z) \cr
&+ \frac{1}{n} \Bigl\{ (n+2) z \omega_+(z)^2 - 2 A_-(z) \omega_+(z) \Bigr\}
\Bigl\{ (n+2) z \omega_-(z) - n t_+ - A_+(z) \Bigr\} \cr
& - \frac{2}{n} K_-(z)  f_+(z)
- \frac{2}{n} K_+(z)  f_-(z) 
+ \frac{4(n+1)}{n} h_-(z)  + 2 g_-(z) = 0.
\end{split}
\ee
Using the planar second order constraint \eqref{PVCr}, we can convert this equation into
an algebraic equation for $\omega_-(z)$:
\bel{cubic2}
\begin{split}
& \frac{4(n-2)}{3} z^2 \, \omega_-(z)^3
+ 4 A_+(z)\, z \omega_-(z)^2 \cr
&- \Bigl\{ A_+(z)^2 - A_-(z)^2 + 4 (n+2) z f_-(z) \Bigr\} \omega_-(z) \cr
& - 2 K_-(z)  f_+(z) + 2 \Bigl\{ (n+2) t_+ + K_+(z) \Bigr\} f_-(z) \cr
& + 4(n+1) h_-(z) + 2n g_-(z) = 0.
\end{split}
\ee
For $n \neq 2$, this is a cubic equation. For $n=2$, it is quadratic.

When $n \neq 2$, in terms of
\be
x_+(z):= (n-2) z \, \omega_-(z) +  A_+(z)
= \frac{(n-2)}{2} v_+(z) + K_+(z),
\ee
the cubic equation \eqref{cubic2} becomes
\bel{cubic2r}
x_+(z)^3 - 3 \, p(z) \, x_+(z) - q_-(z) = 0,
\ee
where $p(z)$ is given by \eqref{pz} and
\be
\begin{split}
q_-(z) &=- \frac{1}{4} A_+(z) 
\Bigl\{ (3n+2) A_+(z)^2 - 3(n-2) A_-(z)^2 \Bigr\} \cr
& + \frac{3(n-2)^2}{2}  z\, K_-(z) f_+(z) 
- \frac{3(n-2)(3n+2)}{2}  z\, K_+(z)  f_-(z) \cr
& - 3(n-2)^2 (n+1) z\, h_-(z) - \frac{3n(n-2)^2}{2} z\, g_-(z).
\end{split}
\ee

\subsubsection{Remark}

For $n \neq -2$, if we introduce $\alpha_-(z)$ and $\beta_-(z)$ as a solution to
\be
\alpha_-(z) \beta_-(z) = p(z), \qq
\alpha_-(z)^3 + \beta_-(z)^3 = q_+(z),
\ee
then
\be
x_-(z) = \alpha_-(z) + \beta_-(z)
\ee
solves the cubic equation \eqref{cubic1r}.

Similarly, for $n \neq 2$, using $\alpha_+(z)$ and $\beta_+(z)$ obeying
\be
\alpha_+(z) \beta_+(z) = p(z), \qq
\alpha_+(z)^3 + \beta_+(z)^3 = q_-(z),
\ee
we have a solution to \eqref{cubic2r}:
\be
x_+(z) = \alpha_+(z) + \beta_+(z).
\ee

\subsection{Loop equations for special cases}

\subsubsection{$n=2$ case}

When $n=2$, \eqref{cubic2} reduces to a quadratic equation
for $2z \, \omega_-(z) = v_+(z)+t_+$:
\be
\begin{split}
& K_+(z) \bigl( v_+(z) + t_+ \bigr)^2
- \frac{1}{2} \Bigl\{ K_+(z)^2 - A_-(z)^2 + 16 z\, f_-(z) \Bigr\} (v_+(z) + t_+) \cr
& - 2 z \, K_-(z) f_+(z) + 2z \bigl\{ 4 t_+ + K_+(z) \bigr\} f_-(z) 
+ 12 z\, h_-(z) + 4 z \, g_-(z) = 0,
\end{split}
\ee
where
\be
A_-(z) = 2 t_- + K_-(z),
\ee
while $x_-(z) = 2 v_-(z) - K_-(z)$ obeys the cubic equation \eqref{cubic1r} with
\be
p(z) = K_+(z)^2,
\ee
\be
\begin{split}
q_+(z) &= A_-(z) \bigl\{ 3 K_+(z)^2 - A_-(z)^2 \bigr\} \cr
& - 24 z \, K_+(z) f_+(z) + 24z\, K_-(z) f_-(z)
-48 z \, h_+(z) + 48 z\, g_+(z).
\end{split}
\ee
Note that in the case of the ABJ(M) matrix model ($k_1 = -k_2$),
\eqref{A_pm} and \eqref{K_pm} imply
\be
 K_+(z) = A_+(z) = \pi \im \kappa_2. 
\ee 
 
\subsubsection{$n=-2$ case}

When $n=-2$, $x_+(z) = -2 v_+(z) + K_+(z)$ obeys the cubic equation \eqref{cubic2r}
with
\be
p(z) = K_-(z)^2,
\ee
\be
\begin{split}
q_-(z) &=A_+(z) \Bigl\{ A_+(z)^2 - 3 K_-(z) \Bigr\}^2 \cr
& + 24 z \, K_-(z) f_+(z) -24 z\, K_+(z) f_-(z)
+ 48 z \, h_-(z) + 48 z\, g_-(z),
\end{split}
\ee
where
\be
A_+(z) = 2 t_+ + K_+(z),
\ee
while \eqref{cubic1} reduces to a quadratic equation for
$2z \,  \omega_+(z) = v_-(z) + t_-$:
\be
\begin{split}
& K_-(z) \bigl( v_-(z) + t_-)^2
+ \frac{1}{2} \bigl\{ A_+(z)^2 - K_-(z)^2 - 16 z\, f_-(z) \bigr\} \bigl( v_-(z) + t_- \bigr) \cr
& - 2z\, K_+(z) f_+(z) + 2 z\, \bigl\{ 4t_- + K_-(z) \bigr\} f_-(z)
+12 z\, h_+(z) - 4z\, g_+(z) = 0.
\end{split}
\ee

\section*{Acknowledgments}
This work was supported by JSPS KAKENHI Grant Number 15K05059. Support from JSPS/RFBR
bilateral collaborations ``Faces of matrix models in quantum field theory and 
statistical mechanics''
(H. I., T. S. and R. Y.) and 
``Exploration of Quantum Geometry via Symmetry and Duality" (T. O.) is
gratefully appreciated. 
The work of T. S. was also supported in part by Fujukai Foundation.

\appendix

\section{Derivation of third order constraint 1: eq. \eqref{HLeq1}}
\label{A}

From \eqref{TOC1} we have
\bel{HL1}
\begin{split}
& 2 \left\langle \sum_i \sum_{j \neq i}
\frac{\ex^{u_i}}{(z-\ex^{u_i})^2} \coth \frac{u_i-u_j}{2} \right\rangle 
-  \left\langle \sum_i \sum_{j \neq i} \frac{1}{(z-\ex^{u_i})} \frac{1}{\sinh^2 \frac{u_i-u_j}{2}}
\right\rangle\cr
&+ 2 \left\langle \sum_i \sum_{j \neq i} \sum_{k \neq i} \frac{1}{z-\ex^{u_i}}
\coth \frac{u_i - u_j}{2} \coth \frac{u_i-u_k}{2} \right\rangle \cr
& - n \left\langle \sum_i \sum_{j \neq i} \sum_a \frac{1}{z- \ex^{u_i}}
\coth \frac{u_i-u_j}{2} \tanh \frac{u_i - w_a}{2} \right\rangle \cr
& - \frac{2 \kappa_1}{g_s} \left\langle
\sum_i \sum_{j \neq i} \frac{u_i}{z - \ex^{u_i}} \coth \frac{u_i-u_j}{2} \right\rangle=0.
\end{split}
\ee
The third term in \eqref{HL1} can be rewritten as
\be
\begin{split}
&  2 \left\langle \sum_i \sum_{j \neq i} \sum_{k \neq i} \frac{1}{z-\ex^{u_i}}
\coth \frac{u_i - u_j}{2} \coth \frac{u_i-u_k}{2} \right\rangle \cr
&= 2 \left\langle \sum_i \sum_{j \neq i} \sum_{k \neq i,j} \frac{1}{z-\ex^{u_i}}
\coth \frac{u_i - u_j}{2} \coth \frac{u_i-u_k}{2} \right\rangle \cr
&+  2 \left\langle \sum_i \sum_{j \neq i}  \frac{1}{z-\ex^{u_i}}
\coth^2 \frac{u_i - u_j}{2} \right\rangle \cr
&= 2 \left\langle \sum_i \sum_{j \neq i} \sum_{k \neq i,j} \frac{1}{z-\ex^{u_i}}
\coth \frac{u_i - u_j}{2} \coth \frac{u_i-u_k}{2} \right\rangle \cr
&+  2 \left\langle \sum_i \sum_{j \neq i}  \frac{1}{z-\ex^{u_i}}
\frac{1}{\sinh^2 \frac{u_i - u_j}{2}} \right\rangle 
+ 2 (N_1-1) \left\langle \sum_i \frac{1}{z-\ex^{u_i}} \right\rangle.
\end{split}
\ee
Here we have used
\be
\coth^2 x = 1 + \frac{1}{\sinh^2 x}.
\ee
Then \eqref{HL1} can be expressed in the form \eqref{HLeq1}  with
\bel{S110}
\begin{split}
S_{11}^{(0)}(z)&:= 2 g_s^3 \left\langle \sum_i \sum_{j \neq i} \sum_{k \neq i,j} \frac{1}{z-\ex^{u_i}}
\coth \frac{u_i - u_j}{2} \coth \frac{u_i-u_k}{2} \right\rangle \cr
&- n \left\langle \widehat{R}^{(3)}_1(z) 
\right\rangle 
- 2 \kappa_1 g_s^2 \left\langle
\sum_i \sum_{j \neq i} \frac{u_i}{z - \ex^{u_i}} \coth \frac{u_i-u_j}{2} \right\rangle,
\end{split}
\ee
\bel{S111}
\begin{split}
S_{11}^{(1)}(z)&:=
2 g_s^2 \left\langle \sum_i \sum_{j \neq i}
\frac{\ex^{u_i}}{(z-\ex^{u_i})^2} \coth \frac{u_i-u_j}{2} \right\rangle \cr
&+ 2 (N_1-1) g_s^2 \left\langle \sum_i \frac{1}{z-\ex^{u_i}} \right\rangle 
+ g_s^2 \left\langle \sum_i \sum_{j \neq i} \frac{1}{(z-\ex^{u_i})} \frac{1}{\sinh^2 \frac{u_i-u_j}{2}}
\right\rangle.
\end{split}
\ee
In the following part, we show that $S_{11}^{(0)}(z)$ \eqref{S110} and $S_{11}^{(1)}(z)$ \eqref{S111} can be 
converted  respectively into \eqref{S110F} and \eqref{S111F}.

\subsection{Rewriting of $S_{11}^{(0)}(z)$: from \eqref{S110} to \eqref{S110F}}

Let us rewrite the first term in the right-handed side of $S_{11}^{(0)}(z)$ \eqref{S110}.
Notice that
\be
\begin{split}
& 2 \sum_i \sum_{j \neq i} \sum_{k \neq i,j}
\frac{1}{z - \ex^{u_i}} \coth \frac{u_i-u_j}{2} \coth \frac{u_i - u_k}{2} \cr
&=\frac{2}{3}  \sum_i \sum_{j \neq i} \sum_{k \neq i,j}
\left( \frac{1}{z - \ex^{u_i}} \coth \frac{u_i-u_j}{2} \coth \frac{u_i - u_k}{2} \right. \cr
& \qq + \left. \frac{1}{z - \ex^{u_j}} \coth \frac{u_j-u_k}{2} \coth \frac{u_j - u_i}{2}
+\frac{1}{z - \ex^{u_k}} \coth \frac{u_k-u_i}{2} \coth \frac{u_k - u_j}{2} \right) \cr
&= \frac{2}{3} \sum_i \sum_{j \neq i} \sum_{k \neq i,j}
\frac{(z^2 + \ex^{u_i+u_j} + \ex^{u_i + u_k} + \ex^{u_j + u_k})}
{(z-\ex^{u_i})(z-\ex^{u_j})(z-\ex^{u_k})} \cr
&= \frac{8z^2}{3} \sum_i \sum_{j \neq i} \sum_{k \neq i,j}
\frac{1}{(z-\ex^{u_i})(z-\ex^{u_j})(z-\ex^{u_k})} \cr
& - \frac{4z}{3} \sum_{i} \sum_{j \neq i } \sum_{k \neq i,j}
\left( \frac{1}{(z-\ex^{u_i})(z-\ex^{u_j}) } + \frac{1}{(z-\ex^{u_i})(z-\ex^{u_k})}
+ \frac{1}{(z-\ex^{u_j})(z-\ex^{u_k})} \right) \cr
& + \frac{2}{3} \sum_i \sum_{j \neq i} \sum_{k \neq i,j}
\left( \frac{1}{z-\ex^{u_i}}+ \frac{1}{z-\ex^{u_j}}
+\frac{1}{z-\ex^{u_k}} \right) \cr
&=\frac{8z^2}{3} \sum_i \sum_{j \neq i} \sum_{k \neq i,j}
\frac{1}{(z-\ex^{u_i})(z-\ex^{u_j})(z-\ex^{u_k})} \cr
& -4 (N_1-2) z \sum_i \sum_{j \neq i} \frac{1}{(z-\ex^{u_i})(z-\ex^{u_j})} 
+ 2 (N_1-1)(N_1-2) \sum_i \frac{1}{z-\ex^{u_i}}.
\end{split}
\ee
Also
\be
\begin{split}
& \sum_i \sum_{j \neq i} \sum_{k \neq i,j} \frac{1}{(z-\ex^{u_i})(z-\ex^{u_j})(z-\ex^{u_k})} \cr
&= \sum_i \sum_j \sum_k \frac{1}{(z-\ex^{u_i})(z-\ex^{u_j})(z-\ex^{u_k})} \cr
& - 3 \sum_i \sum_j \frac{1}{(z-\ex^{u_i})^2 (z- \ex^{u_j})} 
+ 2 \sum_i \frac{1}{(z-\ex^{u_i})^3} \cr
&= \left( \sum_i \frac{1}{z- \ex^{u_i}} \right)^3
+\frac{3}{2} \left[ \left( \sum_i \frac{1}{z-\ex^{u_i}} \right)^2 \right]'
+ \left( \sum_i \frac{1}{z- \ex^{u_i}} \right)''.
\end{split}
\ee
Therefore, 
\be
\begin{split}
& 2 \sum_i \sum_{j \neq i} \sum_{k \neq i,j}
\frac{1}{z - \ex^{u_i}} \coth \frac{u_i-u_j}{2} \coth \frac{u_i - u_k}{2} \cr
&= \frac{8z^2}{3} 
\left\{ \left( \sum_i \frac{1}{z- \ex^{u_i}} \right)^3
+\frac{3}{2} \left[ \left( \sum_i \frac{1}{z-\ex^{u_i}} \right)^2 \right]'
+ \left( \sum_i \frac{1}{z- \ex^{u_i}} \right)'' \right\} \cr
& -4 (N_1-2) z \left[ \left( \sum_i \frac{1}{z-\ex^{u_i}} \right)^2
+ \left( \sum_i \frac{1}{z-\ex^{u_i}} \right)' \right] \cr
& + 2 (N_1-1)(N_1-2) \sum_i \frac{1}{z-\ex^{u_i}}.
\end{split}
\ee
Thus, the first term of $S_{11}^{(0)}(z)$ \eqref{S110} can be finally rewritten as follows
\be
\begin{split}
& 2 g_s^3 \left\langle \sum_i \sum_{j \neq i} \sum_{k \neq i, j}
\frac{1}{z-\ex^{u_i}} \coth \frac{u_i-u_j}{2} \coth \frac{u_i - u_k}{2} \right\rangle \cr
&= \frac{8z^2}{3} \Bigl\langle \hat{\omega}_1(z)^3 \Bigr\rangle 
- 4 (t_1-2g_s) z \Bigl\langle \hat{\omega}_1(z)^2 \Bigr\rangle
+ 2 (t_1 - g_s)(t_1-2g_s) \Bigl\langle \hat{\omega}_1(z) \Bigr\rangle \cr
& + 4z^2 g_s \Bigl\langle \hat{\omega}_1(z)^2 \Bigr\rangle' - 4(t_1-2)z g_s 
\Bigl\langle \hat{\omega}_1(z) \Bigr\rangle' 
+ \frac{8z^2}{3} g_s^2 \Bigl\langle \hat{\omega}_1(z) \Bigr\rangle''.
\end{split}
\ee
This leads to
\bel{S110B}
\begin{split}
S_{11}^{(0)}(z) &= \frac{8z^2}{3} \Bigl\langle \hat{\omega}_1(z)^3 \Bigr\rangle 
- 4 (t_1-2g_s) z \Bigl\langle \hat{\omega}_1(z)^2 \Bigr\rangle
+ 2 (t_1 - g_s)(t_1-2g_s) \Bigl\langle \hat{\omega}_1(z) \Bigr\rangle \cr
& + 4z^2 g_s \Bigl\langle \hat{\omega}_1(z)^2 \Bigr\rangle' - 4(t_1-2 g_s)z g_s 
\Bigl\langle \hat{\omega}_1(z) \Bigr\rangle' 
+ \frac{8z^2}{3} g_s^2 \Bigl\langle \hat{\omega}_1(z) \Bigr\rangle'' \cr
& - n \left\langle \widehat{R}^{(3)}_1(z) \right\rangle
- 2 \kappa_1  g_s^2 \, \log z \,  \left\langle \sum_i \sum_{j \neq i} \frac{1}{z-\ex^{u_i}}
\coth \frac{u_i - u_j}{2} \right\rangle 
+ 2  \left\langle
\widehat{H}_1(z)
\right\rangle.
\end{split}
\ee
By substituting the following identity
\be
\begin{split}
& 2 g_s^2 \sum_i \sum_{j \neq i}
\frac{1}{z- \ex^{u_i}} \coth \frac{u_i - u_j}{2} \cr
&= 2 z \hat{\omega}_1(z)^2 - 2 (t_1 - g_s) \hat{\omega}_1(z)
+ 2 z g_s \hat{\omega}_1(z)'
\end{split}
\ee
 into \eqref{S110B}, we obtain \eqref{S110F}.

\subsection{Rewriting of $S_{11}^{(1)}(z)$: from \eqref{S111} to \eqref{S111F}}

Next, we rewrite $S_{11}^{(1)}(z)$.
The first term in \eqref{S111} can be rewritten as
\be
2 g_s^2 \left\langle \sum_i \sum_{j \neq i}
\frac{\ex^{u_i}}{(z-\ex^{u_i})^2} \coth \frac{u_i-u_j}{2} \right\rangle 
= -  \left(  
\left\langle \sum_i \sum_{j \neq i}
\frac{2 g_s^2\, \ex^{u_i}}{(z-\ex^{u_i})} \coth \frac{u_i-u_j}{2} \right\rangle \right)'.
\ee

Note that
\be
\begin{split}
&\sum_i \sum_{j \neq i}
\frac{2 g_s^2\, \ex^{u_i}}{z-\ex^{u_i}} \coth \frac{u_i - u_j}{2}  \cr
&= g_s^2 \sum_i \sum_{j \neq i}
\left( \frac{\ex^{u_i}}{z- \ex^{u_i}} \frac{\ex^{u_i} + \ex^{u_j}}{\ex^{u_i} - \ex^{u_j}}
+ \frac{\ex^{u_j}}{z - \ex^{u_j}} \frac{\ex^{u_j} + \ex^{u_i}}{\ex^{u_j} - \ex^{u_i}} \right) \cr
&=g_s^2 \sum_i \sum_{j \neq i} \frac{z(\ex^{u_i}+ \ex^{u_j})}{(z-\ex^{u_i})(z-\ex^{u_j})} \cr
&= 2 g_s^2 \sum_i \frac{z \, \ex^{u_i}}{z-\ex^{u_i}} \sum_{j \neq i} \frac{1}{z - \ex^{u_j}} \cr
&= 2 z^2 g_s^2 \left( \sum_i \frac{1}{z- \ex^{u_i}} \right)^2 - 2 (N_1-1) z g_s^2 \sum_i \frac{1}{z-\ex^{u_i}}
+2 z^2 g_s^2 \left( \sum_i \frac{1}{z-\ex^{u_i}} \right)' \cr
&= 2 z^2 \hat{\omega}_1(z)^2 -  2(N_1 -1 ) z g_s \, \hat{\omega}_1(z)
+ 2z^2 g_s \, \hat{\omega}'_1(z).  
\end{split}
\ee
Therefore, the first term in \eqref{S111} can be written by using 
$\hat{\omega}_1(z)$ and its derivatives:
\be
\begin{split}
& 2 g_s^2 \left\langle \sum_i \sum_{j \neq i}
\frac{\ex^{u_i}}{(z-\ex^{u_i})^2} \coth \frac{u_i-u_j}{2} \right\rangle  \cr
&= \left( - 2 z^2 \bigl\langle \hat{\omega}_1(z)^2 \bigr\rangle
+  2(t_1 -g_s ) z  \bigl\langle \hat{\omega}_1(z) \bigr\rangle
- 2z^2 g_s  \bigl\langle \hat{\omega}_1(z) \bigr\rangle'  \right)'.
\end{split}
\ee
Using this relation, we can easily find the final expression \eqref{S111F}.

\section{Derivation of third order constraint 3: eq. \eqref{HLeq3}}
\label{B}

From \eqref{TOC3} we have
\be
\begin{split}
&2 \left\langle \sum_i \sum_a \frac{\ex^{u_i}}{(z-\ex^{u_i})^2} \tanh \frac{u_i - w_a}{2}
\right\rangle 
 + \left\langle \sum_i \sum_a \frac{1}{z- \ex^{u_i}} \frac{1}{\cosh^2 \frac{u_i-w_a}{2}}
\right\rangle \cr
& + 2 \left\langle \sum_i \sum_a \sum_{j \neq i}
\frac{1}{z-\ex^{u_i}} \tanh \frac{u_i - w_a}{2} \coth \frac{u_i-u_j}{2} \right\rangle \cr
&-n \left\langle \sum_i \sum_a \sum_{b}
\frac{1}{z-\ex^{u_i}} \tanh \frac{u_i - w_a}{2} \tanh \frac{u_i-w_b}{2} \right\rangle \cr
& - \frac{2\kappa_1}{g_s} 
\left\langle \sum_i \sum_a \frac{u_i}{z- \ex^{u_i}} \tanh \frac{u_i-w_a}{2} \right\rangle = 0.
\end{split}
\ee
With some work, we can rewrite this constraint in the form \eqref{HLeq3} with
\bel{S120}
\begin{split}
S_{12}^{(0)}(z) &:=
2  \left\langle 
\widehat{R}^{(3)}_1(z)
\right\rangle \cr
& - n g_s^3 \left\langle \sum_i \sum_a \sum_{b \neq a}
\frac{1}{z- \ex^{u_i}} \tanh \frac{u_i - w_a}{2} \tanh \frac{u_i - w_b}{2} \right\rangle \cr
& - 2 \kappa_1 g_s^2 \left\langle \sum_i \sum_a
\frac{u_i}{z-\ex^{u_i}} \tanh \frac{u_i - w_a}{2} \right\rangle,
\end{split}
\ee
\bel{S121}
\begin{split}
S_{12}^{(1)}(z) 
&:= 2 g_s^2 \left\langle \sum_i \sum_a \frac{\ex^{u_i}}{(z-\ex^{u_i})^2} \tanh \frac{u_i - w_a}{2} \right\rangle \cr
& + (n+1) g_s^2 \left\langle \sum_i \sum_a \frac{1}{z-\ex^{u_i}} \frac{1}{\cosh^2 \frac{u_i - w_a}{2}}
\right\rangle
- n t_2 g_s \left\langle \sum_i \frac{1}{z - \ex^{u_i}} \right\rangle.
\end{split}
\ee
Here we have used
\be
\tanh^2 x = 1 - \frac{1}{\cosh^2 x}.
\ee
In the following part, we show that $S_{12}^{(0)}(z)$ \eqref{S120} and $S_{12}^{(1)}(z)$ \eqref{S121} can be 
converted  respectively into \eqref{S120F} and \eqref{S121F}.

\subsection{Rewriting of $S_{12}^{(0)}(z)$: from \eqref{S120} to \eqref{S120F}}

Let us rewrite the second term in $S_{12}^{(0)}(z)$ \eqref{S120}.
We use the following identity
\be
\begin{split}
& \frac{1}{z- \ex^{u_i}} \tanh \frac{u_i - w_a}{2} \tanh \frac{u_i - w_b}{2} \cr
&= \frac{1}{(z-\ex^{u_i})}
\frac{(z-\ex^{w_a})}{(z+ \ex^{w_a})}
\frac{(z-\ex^{w_b})}{( z + \ex^{w_b})} \cr
& - \frac{1}{z+\ex^{w_a}} \tanh \frac{w_a-u_i}{2} \coth \frac{w_a-w_b}{2}
- \frac{1}{z+\ex^{w_b}} \tanh \frac{w_b-u_i}{2} \coth \frac{w_b-w_a}{2} \cr
& - \frac{2z}{(z+\ex^{w_a})(z+\ex^{w_b})}
+ \frac{1}{z+\ex^{w_a}} 
+ \frac{1}{z+\ex^{w_b}}.
\end{split}
\ee
This leads to 
\be
\begin{split}
& g_s^3 \left\langle \sum_i \sum_a \sum_{b \neq a}
\frac{1}{z- \ex^{u_i}} \tanh \frac{u_i - w_a}{2} \tanh \frac{u_i - w_b}{2} \right\rangle \cr
&=g_s^3 \left\langle \sum_i \sum_a \sum_{b \neq a}
\frac{1}{(z-\ex^{u_i})}
\frac{(z-\ex^{w_a})}{(z+ \ex^{w_a})}
\frac{(z-\ex^{w_b})}{( z + \ex^{w_b})} 
\right\rangle + 2  \left\langle \widehat{R}^{(3)}_2(-z) \right\rangle \cr
&-2 t_1 z g_s^2 \left\langle \sum_a \sum_{b \neq a} \frac{1}{(z+\ex^{w_a})(z+\ex^{w_b})}
\right\rangle 
 + 2 t_1 ( t_2-g_s) \Bigl\langle \hat{\omega}_2(-z) \Bigr\rangle \cr
&= \Bigl\langle \hat{\omega}_1(z) \bigl( 2 z \, \hat{\omega}_2(-z) + t_2 \bigr)^2
\Bigr\rangle + 2  \left\langle \widehat{R}^{(3)}_2(-z) \right\rangle 
 - 2 t_1 z \Bigl\langle \hat{\omega}_2(-z)^2 \Bigr\rangle
- 2 t_1 (t_2-g_s) \Bigl\langle \hat{\omega}_2(-z) \Bigr\rangle \cr
&-4 z^2 g_s  \Bigl\langle \hat{\omega}_1(z) \bigl( \hat{\omega}_2(-z)  \bigr)' \Bigr\rangle
- 4 z g_s \Bigl\langle \hat{\omega}_1(z) \hat{\omega}_2(-z) \Bigr\rangle
+ 2 t_1 z  g_s \Bigl\langle \hat{\omega}_2(-z) \Bigr\rangle' - t_2 g_s
\Bigl\langle \hat{\omega}_1(z) \Bigr\rangle.
\end{split}
\ee
Using this relation, \eqref{S120} is easily converted into \eqref{S120F}.
  
\subsection{Rewriting of $S_{12}^{(1)}(z)$: from \eqref{S121} to \eqref{S121F}}

The first term in \eqref{S121} can be rewritten by using the following relation:
\be
g_s^2 \left\langle \sum_i \sum_a \frac{\ex^{u_i}}{(z-\ex^{u_i})^2}
\tanh \frac{u_i - w_a}{2} \right\rangle
= \left( - z \left\langle \widehat{R}_1^{(2)}(z) \right\rangle \right)'.
\ee
With help of this, we can rewrite \eqref{S121} in the form \eqref{S121F}.


\end{document}